\documentclass[12pt]{article}
\usepackage[left=2cm,top=2.5cm,right=2cm,bottom=2.5cm]{geometry}

\usepackage{amsmath,amssymb}

\usepackage{changepage}

\usepackage[utf8x]{inputenc}

\usepackage{textcomp,marvosym}

\usepackage{cite}

\usepackage{nameref,hyperref}

\usepackage[right]{lineno}

\usepackage{microtype}
\DisableLigatures[f]{encoding = *, family = * }

\usepackage[table]{xcolor}

\usepackage{array}
\usepackage{placeins}
\usepackage{subcaption}
\usepackage{authblk}
\usepackage{accents}

\newcolumntype{+}{!{\vrule width 2pt}}

\newlength\savedwidth

\usepackage[aboveskip=1pt,labelfont=bf,labelsep=period,justification=raggedright,singlelinecheck=off]{caption}

\makeatletter
\renewcommand{\@biblabel}[1]{\quad#1.}
\makeatother

\date{}

\usepackage{lastpage,fancyhdr,graphicx}
\usepackage{epstopdf}
\title{Inference of forex and stock-index financial networks based on the normalised mutual information rate}
\author[1]{Yong Kheng Goh\thanks{gohyk@utar.edu.my}}
\author[2]{Haslifah M. Hasim \thanks{hhashim@essex.ac.uk}}
\author[2]{Chris G. Antonopoulos \thanks{canton@essex.ac.uk}}
\affil[1]{Centre for Mathematical Sciences, Universiti Tunku Abdul Rahman, Malaysia}
\affil[2]{Department of Mathematical Sciences, University of Essex, United Kingdom}
\date{}
\begin{document}
\maketitle

\begin{abstract}
In this paper we study data from financial markets using an information-theory tool that we call the normalised Mutual Information Rate and show how to use it to infer the underlying network structure of interrelations in foreign currency exchange rates and stock indices of 14 countries world-wide and the European Union. We first present the mathematical method and discuss about its computational aspects, and then apply it to artificial data from chaotic dynamics and to correlated random variates. Next, we apply the method to infer the network structure of the financial data. Particularly, we study and reveal the interrelations among the various foreign currency exchange rates and stock indices in two separate networks for which we also perform an analysis to identify their structural properties. Our results show that both are small-world networks sharing similar properties but also having distinct differences in terms of assortativity. Finally, the consistent relationships depicted among the 15 economies are further supported by a discussion from the economics view point.
\end{abstract}

\section*{Introduction}
Financial markets can be thought of as complex systems, and are treated as such by the Economics community \cite{plerou2000econophysics, bonanno2001levels, rickles2008econophysics, fiedor2014mutual}. Economists however, approach financial markets from an economics view point, without frequently using information and tools available in other disciplines, such as in Physics and Mathematics \cite{fiedor2014mutual}.

Treating financial markets as a complex system helps in realising the relation to other complex systems and to common approaches in studying them. The performance and interaction of the individual components of such systems are expected to follow a non-linear, chaotic, rather than random, pattern of interaction with each other. It makes thus sense to study the system from the perspective of complex systems theory. Particularly, in this paper we consider data from financial markets world-wide and the European Union (EU). Financial markets consist of components, such as stock prices and currency exchange ratess. The connectivity between these components is usually unknown but highly desirable. The components form a network of nodes and connections, where the nodes are the individual index values, and the connections the interactions among them, usually non-linear and unknown! Here, we use a recently published information-mathematical theory for network inference based on the Mutual Information Rate (MIR) \cite{bianco2016successful} to infer the structure of financial networks coming from recorded data from financial markets. MIR is a measure of the amount of information per unit of time that can be transferred between any two nodes in a network or system \cite{10.1371/journal.pone.0046745,bianco2016successful}.

When network engineers need to infer the topology of a given computer network, they ping individual nodes to map the layout. In financial markets, the analogue to pinging is to using financial time-series data, whereby each component is recorded and tracked regularly in time. We then want to study these data to understand how information flows from node to node in the network and ultimately, its structure, with the links being in general either direct or indirect.

MIR was first introduced by Shannon in 1948 as a ``rate of actual transmission''  \cite{Shannon1948} and, it was redefined more rigorously in \cite{Dobrushin1959} and even later in \cite{Gray1980}. It represents the mutual information (MI) exchanged per unit of time between two dynamical, correlated, variables and is based on mutual information which quantifies linear and non-linear interdependencies between two systems or data sets. It is essentially a measure of how much information two systems or two data sets share. Even though MI is a very important quantity to understand various complex systems ranging from the brain \cite{antonopoulos2015brain} to chaotic systems, there are three main difficulties that need to be overcome: (a) MI in random memoryless processes does not consider the degree of memory that financial markets are proved to contain \cite{fiedor2014mutual,mohammadi2011behaviour,bordignon2008periodic}, (b) it is necessary to determine what is considered as significant event in the complex system under study as probabilities of significant events often need to be known prior to the calculation of MI, (c) due to the usually limited size of data sets, it is complicated to calculate these probabilities accurately. This might lead to a biased calculation of MI \cite{steuer2002mutual}. To overcome these challenges, it has been proposed in \cite{baptista2012mutual,fiedor2014mutual} to calculate the amount of information exchanged per unit of time between two nodes in a dynamical network, aka MIR. This rate permits a more reliable measure of the hierarchical dependency in networks. Other limitations are related to the market structure, that is to sectors and sub-sectors of economic activities for stock markets and geographical locations for market indices and foreign exchange markets. More information can be extracted when studying the causal relationships between markets, either by lead-lag effect (asymmetric correlations) or partial correlations. Another limitation is introduced by the use of linear measures for the relation between market elements, ignoring the complexity in such systems and the fact that financial markets present a non-linear behaviour with regard to stock returns. In \cite{fiedor2014partial, you2015network}, the authors have overcome the above limitations by performing a network analysis of financial markets based on partial mutual information. 

In this paper, we are interested in identifying interdependencies between nodes in the said financial networks and not in inferring the directionality of information flow, i.e. which node is driving which (i.e. causality). We thus assume through out the paper that all networks are undirected, i.e., each connection between two nodes is bidirectional. The existence of such a connection means there is a bidirectional connection between them due to their interaction. The purpose of this work is focused on demonstrating how one can use a quantity based on MIR \cite{bianco2016successful} that we call the normalised MIR to infer the network structure in financial data. Particularly, we show how one can infer the underlying network connectivity among the nodes of the network of financial time-series data, such as foreign currency exchange rates and stock indices. We first apply a normalisation to the MIR using artificial data with given properties, compare the new methodology to work already done in \cite{bianco2016successful} and then apply it to financial market data based on \cite{fiedor2014mutual} for the economies of 14 world-wide countries and the EU. Our results show consistent relationships and structural properties among the 15 economies and, an explanation from the Economics view point is given that further supports these findings.

\section*{Materials and methods}
\subsection*{Information and network connectivity}

As is well-known, a system can produce information which can be transferred among its different components \cite{paluvs1996coarse, schreiber2000measuring, baptista2012mutual, marchiori2012energy, mandal2013maxwell, antonopoulos2014production, antonopoulos2015brain}. When information is transferred, there are at least two components involved that are physically interacting by either direct or indirect ways. These components can be either time-series data, modes, or related functions, and from the mathematical view point, they are defined on subspaces or projections of the state space of the system \cite{baptista2012mutual, antonopoulos2014production}. In this work, we will use a quantity based on MIR, the normalised MIR, to study the amount of information transferred per unit of time between any two components of a system, namely to determine whether a link between these components exists. Such an existence means there is an undirected connection between them attributed to their interaction.

MIR measures the rate of information exchanged per unit of time between any two non-random, correlated variables. Since variables in complex systems are not purely random, MIR is an appropriate quantity to measure the amount of information exchanged per unit of time among nodes in such systems. Its application to time-series data and to their functions is of primordial importance to our work and will be used to determine if an (undirected) link exists between any two nodes in the system. In this framework, the strength of such connections can also be inferred, in the sense that they will be compared to those from all other pairs of nodes in the network, so long as the available data collected from the financial markets are sufficient to allow for a discrimination between stronger and weaker connections.

\subsection*{Mutual information}

MI and MIR  were originally introduced by Shannon in 1948 \cite{Shannon1948}. Particularly, the MI of two random variables $X$ and $Y$ is a measure of their mutual dependence and quantifies the ``amount of information'' obtained about one random variable $X (Y)$, after observing the other random variable $Y (X)$. It is defined by \cite{Shannon1948,kullback1997}

\begin{equation}\label{IXY1}
I_{XY}\left ( N \right )=H_{X}+H_{Y}-H_{XY},
\end{equation}
where $N$ is the total number of random events in variables $X$ and $Y$. $H_{X}$ and $H_{Y}$ are the marginal entropies of $X$ and $Y$ (i.e. the Shannon entropies) respectively, defined by
\begin{equation} H_{X}=-\sum_{i=1}^{N}P_{X}(i) \log P_{X}(i) \end{equation}
and 
\begin{equation} H_{Y}=-\sum_{j=1}^{N}P_{Y}(j) \log P_{Y}(j), \end{equation}
where $P_{X}(i)$ is the probability of a random event $i$ happening in $X$, and $P_{Y}(j)$ is the probability of a random event $j$ happening in $Y$. 

The joint entropy, $H_{XY}$ in Eq. \eqref{IXY1} measures how much uncertainty there is in $X$ and $Y$ when taken together. It is defined as
\begin{equation}\label{HXY}
H_{XY}=-\sum_{i=1}^{N}\sum_{j=1}^{N}P_{XY}(i,j) \log P_{XY}(i,j),
\end{equation}
where $P_{XY}(i,j)$ is the joint probability of both events $i$ and $j$ occurring simultaneously in variables $X$ and $Y$.

Equivalently, we can define MI by
\begin{equation}
I_{XY}\left ( N \right )=\sum_{i=1}^{N}\sum_{j=1}^{N}P_{XY}\left ( i,j \right ) \log \left ( \frac{P_{XY}\left ( i,j \right )}{P_{X}(i)P_{Y}(i)} \right ).\label{IXY2}
\end{equation}
This equation provides a measure of the strength of the dependence between the two random variables, and the amount of information $X$ contains about $Y$, and vice versa \cite{kullback1997}. When MI is zero, $I_{XY}=0$, the strength of the dependence is null and thus $X$ and $Y$ are independent: knowing $X$ does not give any information about $Y$ and vice-versa. Note that MI is also symmetric, i.e. $I_{XY}=I_{YX}$.

The computation of $I_{XY}\left (N \right)$ from time-series data requires the calculation of probabilities in an appropriately defined probabilistic space on which a partition can be defined (see for example Eqs. \eqref{HXY} and \eqref{IXY2}). The probabilistic space is defined based on the $X$, $Y$ variables. $I_{XY}$ can be computed for any two components, or nodes $X$ and $Y$ of the same network and can then be compared with $I_{XY}$ of any other pair of nodes in the same network. However, $I_{XY}$ is not suited to compare between different systems as it is possible for different systems to have different correlation decay-times and time-scales \cite{eckmann1985ergodic, baptista2008finding, pinto2011density} in their dynamical evolution.

There are various methods to compute MI, depending on the method used to calculate the probabilities in Eq. \eqref{IXY2}. The main methods are the binning (histogram) method \cite{moddemeijer1989estimation}, the density-kernel method \cite{moon1995estimation}, and the estimation of probabilities from distances between closest neighbours \cite{kraskov2004estimating}. In this work, we use the binning (histogram) method \cite{bianco2016successful}. Binning tends to overestimate MI because of the finite length of recorded time-series data, and the finite resolution of non-Markovian partitions \cite{butte2000mutual, steuer2002mutual}. However, these errors are systematic and always present for any given non-Markovian partition. To avoid such errors, we apply a normalisation as proposed in \cite{bianco2016successful} when dealing with MIR, utilising the binning method to calculate the probabilities in Eq. \eqref{IXY2}.

\subsection*{Mutual Information Rate}

The Mutual Information Rate came about as a method to bypass problems associated with the resolution of non-Markovian partitions, specifically in calculating MI for such partitions. In \cite{baptista2012mutual}, it was shown how to calculate MIR for two finite length time-series irrespective of the partitions in the probabilistic space. MIR is invariant with respect to the resolution of the partition and is defined by
\begin{align}\label{MIR1}
\mbox{MIR}_{XY} &=\lim_{N\to\infty} \lim_{L\to\infty} \sum_{i=1}^{L-1} \frac{I_{XY}\left ( i+1,N \right )-I_{XY}\left ( i,N \right )}{L} \nonumber \\[1em]
  &=\lim_{N\to\infty} \lim_{L\to\infty}\frac{I_{XY}\left ( L,N \right )-I_{XY}\left ( 1,N \right )}{L} \nonumber \\[1em]
  &=\lim_{N\to\infty} \lim_{L\to\infty}\frac{I_{XY}\left ( L,N \right )}{L}, 
\end{align}
where $I_{XY}\left ( L,N \right )$ represents the MI of Eq. \eqref{IXY1} between two random variables $X$ and $Y$, considering trajectories of length $L$ that follow an itinerary over cells in an $N\times N$ grid of equally sized boxes defined in the probabilistic space. Note that the term $I_{XY}\left (1,N \right)/{L}$ tends to zero in the limit of infinitely long trajectories, (i.e. $L\rightarrow \infty$).

For finite-length time-series $X$ and $Y$, the definition in Eq. (6) can be further reduced, as demonstrated in \cite{baptista2012mutual}, to
\begin{equation}\label{MIR_def2}
\mbox{MIR}_{XY}=\frac{I_{XY}\left ( N \right )}{T\left ( N \right )}, \end{equation}
where $I_{XY}\left ( N \right )$ is defined as in Eqs. \eqref{IXY1} and \eqref{IXY2}, and $N$ is the total number of cells in a Markov partition of order $T$ for a particular grid-size $N$.

It is important to note that while $T$ and $N$ are both finite quantities, for statistically significant results, a sufficiently large number of points is required to ensure that the length of the time-series is sufficiently larger than $T$, and thus a more saturated distribution of data across the probabilistic space and its partitions can be achieved.

\section*{Demonstration of the method}

We first demonstrate the usefulness of the method to infer the structure of networks where the dynamics in each node is given by a chaotic map, before applying it to financial-markets data. Particularly, we apply the proposed methodology to networks with known structure given by their adjacency matrices, and different maps for the dynamics of their nodes, and assess the obtained results by comparing the inferred network with the original networks, i.e. by comparing their adjacency matrices to decide whether the inference was successful in terms of the number of correctly inferred connections expressed as a percentage. For example, 100\% successful inference means that the proposed methodology was able to infer all connections in the original network used to produce the recorded data with no spurious connections appearing. 

We first generate time-series data using discrete-time dynamics for the nodes of networks with given binary adjacency matrices. For these data and for each considered system, we then calculate the normalised MIR for each pair of nodes in the network, and subsequently its adjacency matrix based on the computed values and an appropriately defined inference threshold. Finally, we compare the adjacency matrix of the inferred network with that of the initially used network to estimate the percentage of successful inference.

\subsection*{Circle map network}

\begin{figure}[!ht]
\centering\includegraphics[width=0.8\textwidth]{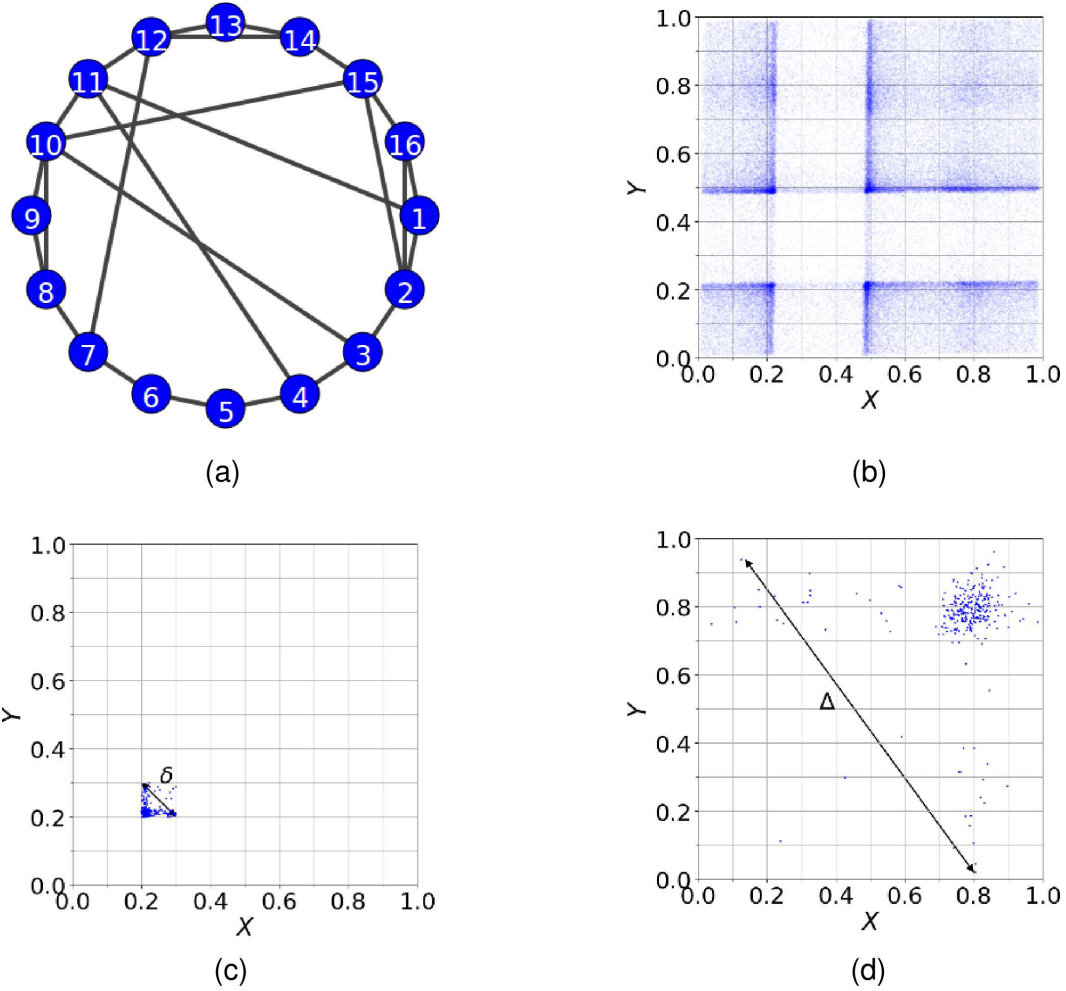}
\caption{\textbf{The CMN, distribution of data points and expansion of points in $\Omega$.} Panel (a): The CMN is composed of 16 coupled nodes as shown by its network. The dynamics in each node is given by Eqs. \eqref{CMN} and \eqref{dynamics_coupling_equation}. Panel (b): The distribution of points in $\Omega$ obtained from Eqs. \eqref{CMN} and \eqref{dynamics_coupling_equation}, plotted in a $10 \times 10$ grid of equally sized cells. Panel (c): The points belong initially to a cell of the same grid and expand to a larger portion of $\Omega$ after three iterations, occupying more than one cells. $\delta$ is the maximum distance in the initial cell and $\Delta$ the maximum distance after the points have expanded to a larger extend of $\Omega$.}
\label{CMN_fig}
\end{figure}

Following \cite{bianco2016successful}, we first apply the methodology to the case of a topology, which we refer to as the circle map network (CMN) (see Fig~\ref{CMN_fig}a). The network is composed of 16 coupled nodes with dynamics given by \cite{kaneko1990clustering}
\begin{equation}\label{CMN}
x_{n+1}^{i}=\left ( 1-\alpha \right )f\left ( x_{n}^{i},r \right )+\frac{\alpha }{k_{i}} \sum_{j=1}^{M} A_{ij} f\left ( x_{n}^{j},r \right ),
\end{equation}
where $M=16$ is the total number of nodes, $x_{n}^{i}$ is the $n$-th iterate of map $i$ (where $i=1,2,\ldots,M$) and $\alpha \in \left [ 0,1 \right ]$ is the coupling strength. $(A_{ij})$ is the binary adjacency matrix of connectivities, containing entries of either 1 when there is a connection between nodes $i$ and $j$, or 0 when there is no such connection. Moreover, it is a symmetric matrix as we consider a connection between any two nodes to be undirected (i.e. bidirectional). $k_{i}=\displaystyle \sum_{j=1}^{M} A_{ij}$ is the node-degree, $r$ the parameter of each map, and $f\left (x_{n},r \right)$ the circle map, defined by
\begin{equation}\label{dynamics_coupling_equation}
f\left ( x_{n},r \right )=x_{n}+r-\frac{K}{2 \pi} \sin\left ( 2\pi x_{n} \right ) \mod 1.
\end{equation}

For the parameters in Eq. \eqref{CMN}, we follow \cite{bianco2016successful} and use $\alpha=0.03$ to create a weakly interacting system, and $r=0.35$ and $K=6.9115$ that correspond to fully developed chaos for each individiual map. $x_{0}^{i}$ is initialised randomly, and the first 1,000 iterations are discarded so that a random seed can be used to start the next 100,000 iterations generated from the dynamics in the network, which we recorded as the time-series data. Fig~\ref{CMN_fig}a shows the network topology described by the adjacency matrix $A$ with the 16 nodes.

\FloatBarrier

We then calculate MI and MIR for a pair of nodes $X,Y$ by defining a 2-dimensional probability space $\Omega$ formed by the time-series data recorded for that pair (see Fig~\ref{CMN_fig}b). $\Omega$ is partitioned into a grid of $N\times N$ fixed-sized cells where the probability of an event $i$ in $X$ is
\begin{equation} P_{X}\left ( i \right )=\frac{\mbox{number of data points in column } i}{\mbox{total number of data points in } \Omega}, \end{equation}
and the probability of an event $j$ in $Y$ is
\begin{equation} P_{Y}\left ( j \right )=\frac{\mbox{number of data points in column } j}{\mbox{total number of data points in } \Omega}. \end{equation}
Similarly, the joint probability can be defined by the ratio of points in cell $\left ( i,j \right )$ of the same partition in $\Omega$ and is expressed by
\begin{equation}
P_{XY}\left (i,j \right )=\frac{\mbox{number of data points in cell } \left ( i,j \right )}{\mbox{total number of data points in } \Omega}.
\end{equation}
MI therefore can be calculated from Eq. \eqref{IXY2} for different values of grid sizes $N$.

To ensure there is always a sufficiently large number of data points in the cells of the $N \times N$ partition of $\Omega$, we require the average number of points in all occupied cells to be sufficiently larger than the number of occupied cells,
\begin{equation}
\left \langle N_{0}\left ( N \right ) \right \rangle \geq N_{oc},
\end{equation}
where $N_{oc}$ is the number of occupied cells and $\left \langle N_{0}\left ( N \right ) \right \rangle$ is the average number of points in all occupied cells in $\Omega$. For the CMN, we have used a data set of 100,000 points which guarantee that this condition is satisfied for grid sizes up to $N_{\max}=19$ and, thus we have calculated $I_{X,Y}$ for grid sizes ranging from $0.2N_{\max}$ to $N_{\max}$.

In order to compute MIR, as in Eq. \eqref{MIR_def2}, to infer the topology of the network, we first need to estimate $T(N)$, i.e. the correlation decay time. $T(N)$ can be calculated in many ways, e.g. by using the diameter of an associated itinerary graph $G$ \cite{bianco2016successful}, or the Lyapunov exponents of the dynamics or the largest expansion rates \cite{baptista2012mutual}. All these different ways of estimating $T(N)$ exploit the fact that the dynamics is chaotic and uses thus the assumption that the points expand to the whole extend of $\Omega$ after about $T(N)$ time.

For the itinerary graph $G$, each cell in $\Omega$ occupied by at least one point is represented as a node in $G$. If $G_{ij}=1$, there is at least one point that can move from node $i$ to node $j$. If $G_{ij}=0$, then there is no such point. The diameter of $G$ is the minimum distance (hops) required to cover the entire network. The correlation decay time, $T(N)$, is then the minimum time for points in any cell of $\Omega$ to spread to its whole extend, and can be expressed as the diameter of $G$. A drawback of this approach however is that the resulting correlation-decay time is integer and thus not as good approximation as the correlation-decay time estimated by the largest Lyapunov exponent or expansion rate, with profound consequences in the calculation of MIR.

Consequently, in this work, we estimate $T(N)$ by the largest expansion rate which is computed from the recorded data. Particularly,
\begin{equation}\label{TN}
T\left ( N \right )\approx \frac{1}{\lambda _{1}} \log \frac{1}{N},
\end{equation}
where $\lambda _{1}$ is the largest positive Lyapunov exponent of the dynamics \cite{Benettin1980}. However, in a system or for recorded data sets for which $\lambda _{1}$ cannot be estimated or computed, it can be replaced by the largest expansion rate $e_{1}$, defined by
\begin{equation}\label{e_1}
e_{1}=\frac{1}{N_{oc}}\sum_{i=1}^{N_{oc}} \frac{1}{t} \log L_{1}^{i}\left ( t \right ),
\end{equation}
with $e_1\leq\lambda_1$ in general. The equality holds when the system has constant Jacobian, is uniformly hyperbolic, and has a constant natural measure \cite{baptista2012mutual}.

In Eq. \eqref{e_1}, $L_{1}^{i}\left ( t \right )$ is the largest distance between pairs of points in cell $i$ at time $t$ divided by the largest distance between pairs of points in cell $i$ at time 0, and is expressed as
\begin{align} 
L_{1}^{i}\left ( t \right ) &=\frac{\Delta }{\delta } \nonumber\\[1em]
    &=\frac{\mbox{largest distance between pairs of point in cell } i \mbox{ at time }t}{\mbox{largest distance between pairs of point in cell } i \mbox{ at time }0} . 
\end{align}

In Fig~\ref{CMN_fig}c and Fig~\ref{CMN_fig}d we present an example of the expansion of points for the CMN that initially belong to a single cell (Fig~\ref{CMN_fig}c) and after three iterations of the dynamics, they expand to a larger portion of $\Omega$ (Fig~\ref{CMN_fig}d). In Fig~\ref{CMN_fig}c,  we denote by $\delta$ the maximum distance for a pair of points in the initial cell and in Fig~\ref{CMN_fig}d by $\Delta$ the maximum distance after they have expanded to a larger extend of $\Omega$.

In the calculation of MIR, since we are using partitions of fixed-size cells which are non-Markovian, errors will populate, causing a systematically biased computation towards larger MIR values. In such cases, MIR is partition-dependent, and we thus calculate it in a way that makes it partition-independent, following \cite{bianco2016successful}. Particularly, there is a systematic error coming from the non-Markovian nature of the equally-sized cells in the considered grids as a smaller $N$ is likely to create a partition which is significantly different to a Markovian one than a larger grid size $N$. Moreover, using the fact that $\mbox{MIR}_{XY}=\mbox{MIR}_{YX}$ and $\mbox{MIR}_{XX}=0$, we can narrow the number of $X$, $Y$ pairs from $M^2$ to $M\left ( M-1 \right )/{2}$. The method presented in \cite{bianco2016successful} to avoid such errors is to normalise MIR for each grid size $N$ as follows: For fixed $N$, $\mbox{MIR}_{XY}\left ( N \right )$ is first computed for all $M(M-1)/2$ pairs of nodes and is normalised with respect to their minimum and maximum values. The reason is that for unconnected pairs of nodes, MIR is numerically very close to zero, and doing so the new $\widehat{\mbox{MIR}}_{XY}$ in Eq. \eqref{MIR_wedge} will be in the interval $[0,1]$. This normalisation can be achieved by
\begin{equation}\label{MIR_wedge}
\widehat{\mbox{MIR}}_{XY}\left ( N \right )=\frac{\mbox{MIR}_{XY}\left ( N \right )-\min\left \{ \mbox{MIR}_{XY}\left ( N \right ) \right \}}{\max \left \{ \mbox{MIR}_{XY}\left ( N \right )\right \}-\min \left \{ \mbox{MIR}_{XY} \left ( N \right )\right \}},
\end{equation}
where $\mbox{MIR}_{XY}\left ( N \right )$ is the MIR calculated for nodes $X$ and $Y$, $\min \left \{ \mbox{MIR}_{XY} \left ( N \right ) \right \}$ is the minimum MIR of all $M(M-1)/2$ pairs, and similarly $\max \left \{ \mbox{MIR}_{XY} \left ( N \right ) \right \}$ is the maximum MIR over all MIR values of the $M(M-1)/2$ pairs. Since we use $\widehat{\mbox{MIR}}_{XY}\left ( N \right )$ for a range of grid sizes $N$, we can further normalise $\widehat{\mbox{MIR}}_{XY}$ by \cite{bianco2016successful}
\begin{equation}\label{MIR_bar}
\overline{\mbox{MIR}}_{XY}=\frac{\sum_{i} \widehat{\mbox{MIR}}_{XY}\left ( N_{i} \right ) }{\max \left \{ \sum_{i} \widehat{\mbox{MIR}}_{XY}\left ( N_{i} \right ) \right \} },
\end{equation}
where the maximum is now taken over the $N_{i}$ grid sizes considered in $[0.2N_{\max}$, $N_{\max}]$. This normalisation ensures again that $\overline{\mbox{MIR}}_{XY}$ values are in $[0,1]$.

To reconstruct a network using Eq. \eqref{MIR_bar}, we fix a threshold $\tau \in [0,1]$ and consider the pair $XY$ as connected if $\overline{\mbox{MIR}}_{XY} \geq \tau$. If so, then the corresponding entry in the adjacency matrix $A^{c}$ of the reconstructed network becomes 1, i.e. $A_{XY}^{c}=1$ or 0 if $\overline{\mbox{MIR}}_{XY} < \tau$ as the pair is considered as unconnected.

\begin{figure}[!h]
  \centering
  \includegraphics[width=0.8\textwidth]{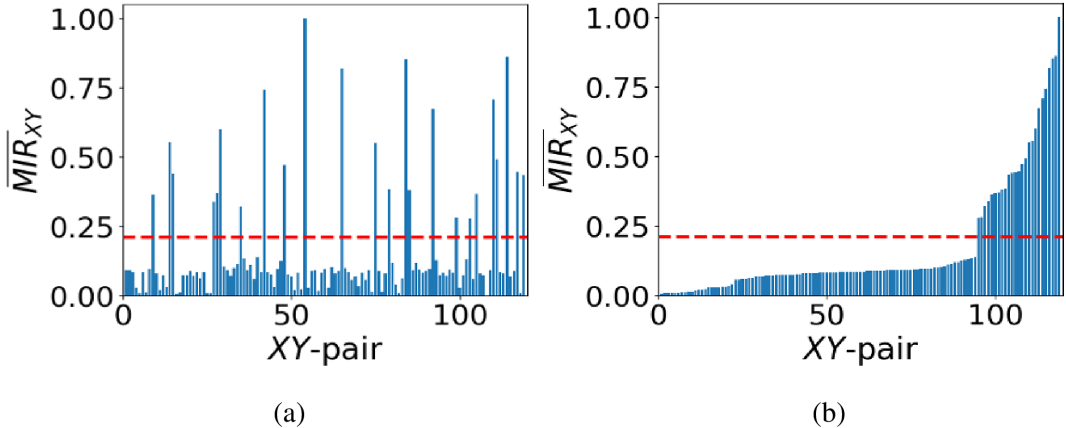}
\caption{\textbf{Estimation of the threshold $\mathbb{\tau}$ for the inference of CMN.} Panel (a) is the plot of the $\overline{\mbox{MIR}}_{XY}$ values before ordering in ascending order and (b) after being ordered. Following the approach in the text, $\tau\approx0.21$ plotted by the red dash line in both panels.}\label{MIR_bar_CNM_fig}
\end{figure}

Choosing $\tau$ appropriately is crucial in successfully depicting the structure of the original network from the recorded data. If $\tau$ is set too high, real connections among nodes might be missed, while if set too low, spurious connections between nodes might appear in the inferred network. To address this, we follow \cite{rubido2014exact} and determine $\tau$ by first sorting all $\overline{\mbox{MIR}}_{XY}$ values in ascending order, and then identifying the first $XY$ pair for which $\overline{\mbox{MIR}}_{XY}$ increases more than 0.1 - which accounts for an abrupt change in $\overline{\mbox{MIR}}_{XY}$ values - and then set the threshold $\tau$ as the middle value of the $\overline{\mbox{MIR}}_{XY}$ for the identified pair and that for the immediately previous pair. This is based on the observation that there two main groups of $\overline{\mbox{MIR}}_{XY}$ values, i.e. for the connected and unconnected nodes (see Fig \ref{MIR_bar_CNM_fig}b).

\FloatBarrier

We demonstrate the application of this approach for network inference in Fig~\ref{MIR_bar_CNM_fig}. Particularly, in panel (a) we plot the $\overline{\mbox{MIR}}_{XY}$ values for all pairs $XY$ before ordering them in ascending order and in (b) the same values ordered. It is evident in panel (b) that there two groups of $\overline{\mbox{MIR}}_{XY}$ values, namely one for the unconnected pairs (left side of the plot with relatively low values) and another one for the group of connected pairs (right side of the plot with relatively big value), separated by an abrupt change in $\overline{\mbox{MIR}}_{XY}$ values in between. Following the proposed methodology, we have estimated that $\tau\approx0.21$ (plotted by the red dash line in both panels) which crosses the bar for which there is this abrupt change. This then leads to the 100\% successful inference of the original network shown in Fig~\ref{CMN_fig}a.

\section*{Inferring networks using data from additional sources}

\subsection*{Introducing an additional node with uniform, uncorrelated, random data}

One of the main difficulties in inferring the structure of a network is the estimation of $\tau$. The notion of an abrupt change might be subjective (more than one abrupt changes) or even not clear-cut, especially when dealing with data from real experiments and/or when the number of nodes in the network is big. Here, we introduce another approach to improve network inference by setting $\tau$ according to the $\overline{\mbox{MIR}}_{XY}$ value of a pair of, disconnected from the network, nodes with data with known properties. This $\overline{\mbox{MIR}}_{XY}$ value will then be used to set the threshold for network inference.

\begin{figure}[!ht]
 \centering
  \includegraphics[width=0.8\textwidth]{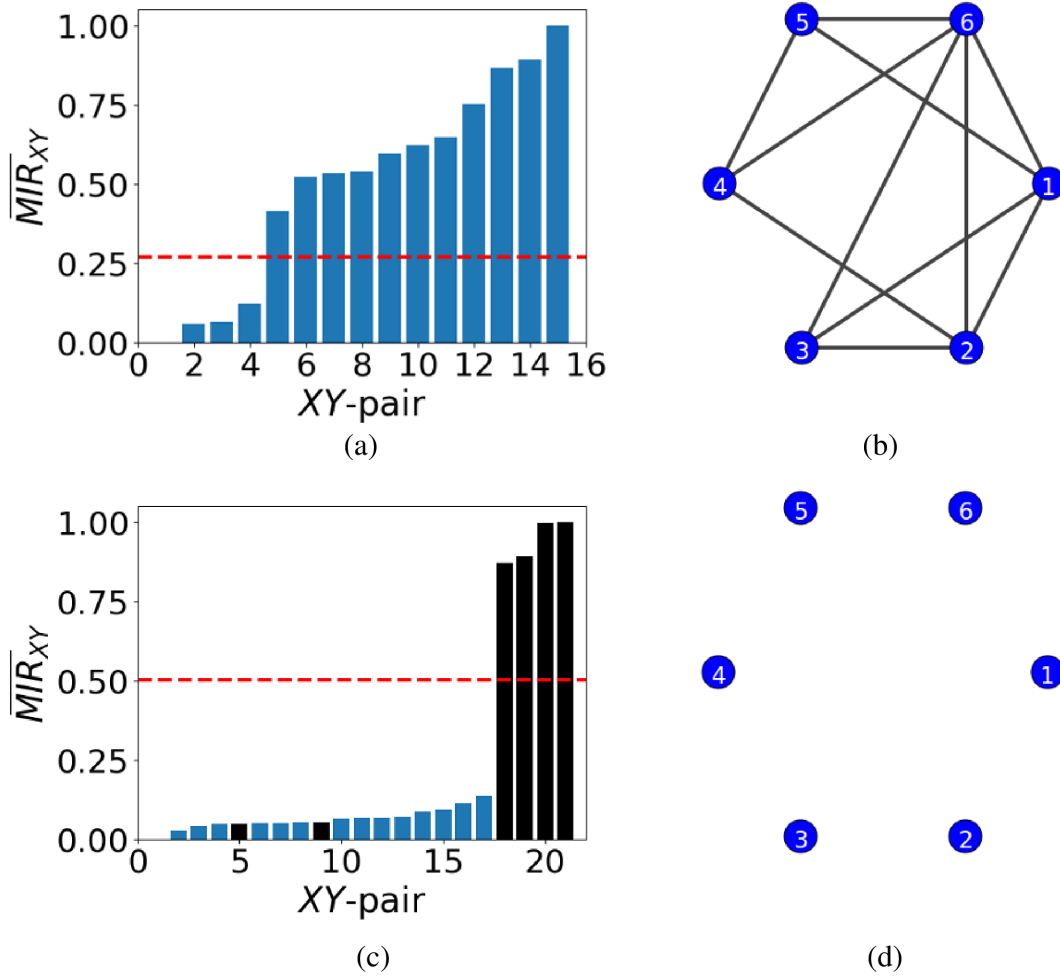}
\caption{\textbf{Results for the case of an additional source of data from a pair of nodes with uniformly random, uncorrelated data.} Panel (a) shows the estimation of $\tau\approx0.27$ based on the ordered $\overline{\mbox{MIR}}_{XY}$ values for a network of 6 isolated nodes without the introduction of the pair of nodes with uniformly random, uncorrelated data. Panel (b) is the unsuccessfully inferred network based on panel (a). Panel (c) shows the ordered $\overline{\mbox{MIR}}_{XY}$ values for the same network with the nodes of random data added. The black bars are the $\overline{\mbox{MIR}}_{XY}$ values that come from the pair of additional nodes. Panel (d) shows the resulting successfully inferred network of isolated, disconnected nodes. In panels (a) and (c), we plot $\tau$ by a red dash line where $\tau\approx0.27$ in (a) and $\tau\approx 0.5$ in (c). 
}\label{fig:rand_ref}
\end{figure}

To demonstrate this, we will use uniformly random, uncorrelated, data as the additional source. One would expect that the $\overline{\mbox{MIR}}_{XY}$ and correlation decay time, $T(N)$, between any two nodes of a network of such data would be close to zero and one, respectively. This idea can be appreciated by using a network of isolated nodes. To this end, we set $\alpha = 0$ in Eq.~(\ref{CMN}), $N=6$ nodes and $f(x_n,r)$ be the logistic map in its chaotic regime (i.e. $r=4$),
\begin{equation}\label{LMAP}
f_{i}\left ( x_{n}^{i},r \right )=rx_{n}^{i}\left ( 1-x_{n}^{i} \right ).
\end{equation}
The normalisation process casts the $\overline{\mbox{MIR}}_{XY}$ values in $[0,1]$ for all pairs of nodes $XY$. Often, the resulting ordered $\overline{\mbox{MIR}}_{XY}$ values do not allow for a clear determination of $\tau$ as evidenced in Fig~\ref{fig:rand_ref}a. Without the use of an additional source of uniformly random, uncorrelated data for a pair of nodes disconnected from the rest of the network, one would compute $\tau\approx0.27$ and would result in the inferred network in panel (b) which is clearly not the original one which consists of only isolated, disconnected nodes! In contrast, when using the idea of introducing uniformly random, uncorrelated data for a pair of nodes disconnected from the rest of the network, and applying the same methodology, panel (c) shows an abrupt change in the ordered $\overline{\mbox{MIR}}_{XY}$ values that can be exploited to infer the network structure. In this case, $\tau\approx0.5$ and leads to the successful network inference shown in panel (d). The black bars correspond to the $\overline{\mbox{MIR}}_{XY}$ values for the pair of introduced nodes with other nodes in the network.

\FloatBarrier

\subsection*{Indirect information exchange and bidirectional connections}

\begin{figure}[!ht]
 \centering 
  \includegraphics[width=0.8\textwidth]{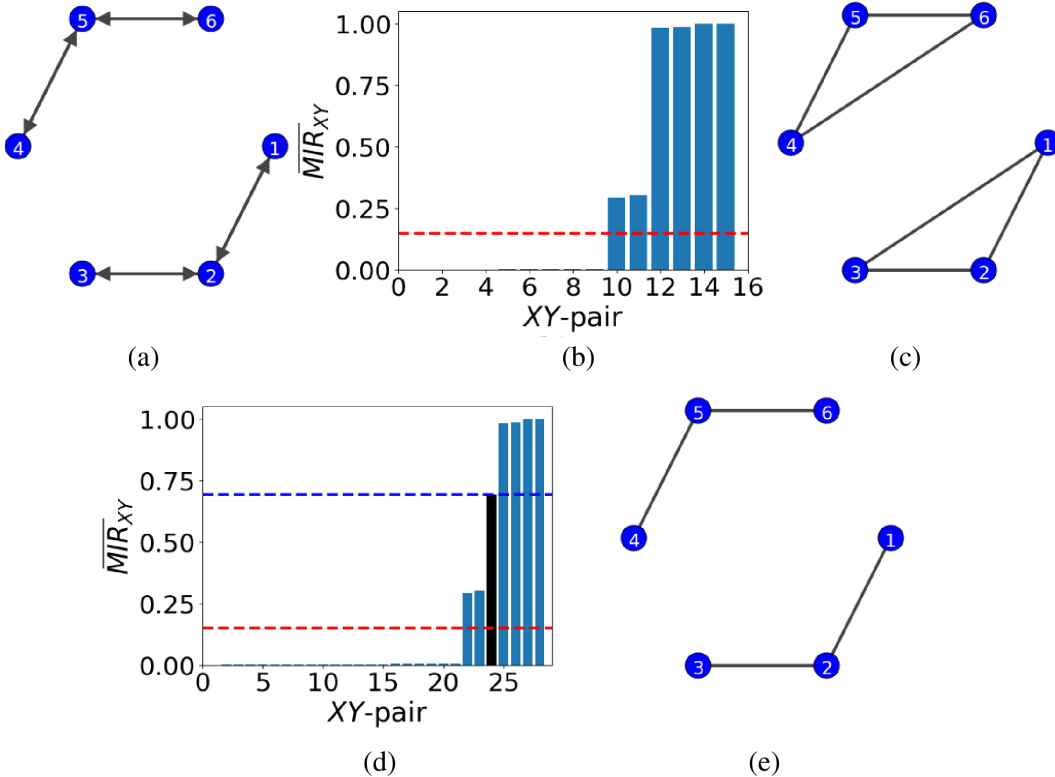}
\caption{\textbf{Network inference in the case of two, disconencted from each other, triplets of nodes.} Panel (a) shows the network of weakly coupled ($\alpha=0.1$) logistic maps. Nodes 1 and 3 interact indirectly through node 2, and similarly, nodes 4 and 6 through node 5 only. Panel (b) shows the ordered $\overline{\mbox{MIR}}_{XY}$ and threshold $\tau$ (red dash line) when no additional nodes are introduced to the network. The red dash line corresponds to $\tau\approx0.16$. Panel (c) shows the unsuccessfully inferred network when using only the data from the network in panel (a). Panel (d) shows the ordered $\overline{\mbox{MIR}}_{XY}$ values for the same network as in (a) with the additional data from the pair of directed nodes (see text). The blue dash line represents the threshold computed for the pair of directed nodes ($\tau\approx0.69$) and the red dash line corresponds to $\tau\approx0.16$ from panel (b). Finally, panel (e) shows the successfully inferred network by considering as connected nodes only those with $\overline{\mbox{MIR}}_{XY}$ bigger than the blue dash threshold. 
} \label{logmap_fig}
\end{figure}

Next, we will examine a network of six weakly coupled ($\alpha=0.1$) logistic maps \eqref{LMAP} seen in Fig~\ref{logmap_fig}a that corresponds to the following adjacency matrix
$$
\left(A_{ij}\right) = \begin{bmatrix} 
0&1&0&0&0&0\\
1&0&1&0&0&0\\
0&1&0&0&0&0\\
0&0&0&0&1&0\\
0&0&0&1&0&1\\
0&0&0&0&1&0
\end{bmatrix}.
$$
This network consists of two triplets of nodes which are disconnected from each other. In each triplet, the two end nodes do not exchange information directly, but only indirectly through the intermediate nodes (i.e. 2 and 5, respectively). Applying the proposed methodology for network inference without the use of any additional nodes will result in picking up the indirect exchange of information as a direct one, where {$\tau\approx0.16$}, depicted as the red dash line in Fig~\ref{logmap_fig}b, leading thus to a spurious direct connection between the end point nodes (see Fig~\ref{logmap_fig}c).

\FloatBarrier

The addition of a new pair of nodes with uniformly random, uncorrelated data disconnected from the rest of the network will not help either in inferring the network structure successfully in this case as the chaotic dynamics data for which we want to infer the network structure are also uncorrelated in time. An alternative approach is  to add a pair of directed nodes to the network, again disconnected from the main network. These nodes will be represented by chaotic logistic maps (i.e. with $r=4$ and $\alpha=0.1$ in Eq. \eqref{LMAP}) with the adjacency matrix
$$
\left(A_{ij}\right) = \begin{bmatrix} 
 0&1\\
 0&0
\end{bmatrix}.
$$
Only the second node is coupled to the first and, thus, the information exchange is unidirectional from the second to the first, and the $\overline{\mbox{MIR}}_{XY}$ would not be as high as for a bidirectional connection between them. Consequently, we may assume that any information exchange value smaller than the $\overline{\mbox{MIR}}_{XY}$ value for this particular pair will not be considered as a connection and will be represented by 0 in the corresponding inferred adjacency matrix (see Fig~\ref{logmap_fig}d where the new threshold, depicted as the blue dash line, is now set at $\tau\approx0.69$ and the old one is represented by the red dash line at $\tau\approx0.16$). Following the proposed methodology for network inference for the augmented data and for $\tau\approx0.69$, we arrive at the  100\% successfully inferred network in Fig~\ref{logmap_fig}e which is the same as in Fig~\ref{logmap_fig}a.

\subsection*{Application to correlated normal variates data}

An interesting application of the proposed methodology is to correlated normal variates data. In most cases, data collected from real-world applications have no obvious dynamical system equations to help relate the different variables involved. For example, global financial markets of different countries are correlated, however the underlying dynamics are not known \cite{junior2015dependency}. The proposed method would be able to provide clues on the interrelations among the financial markets.

\begin{figure}[!ht]
  \centering
 \includegraphics[width=0.8\textwidth]{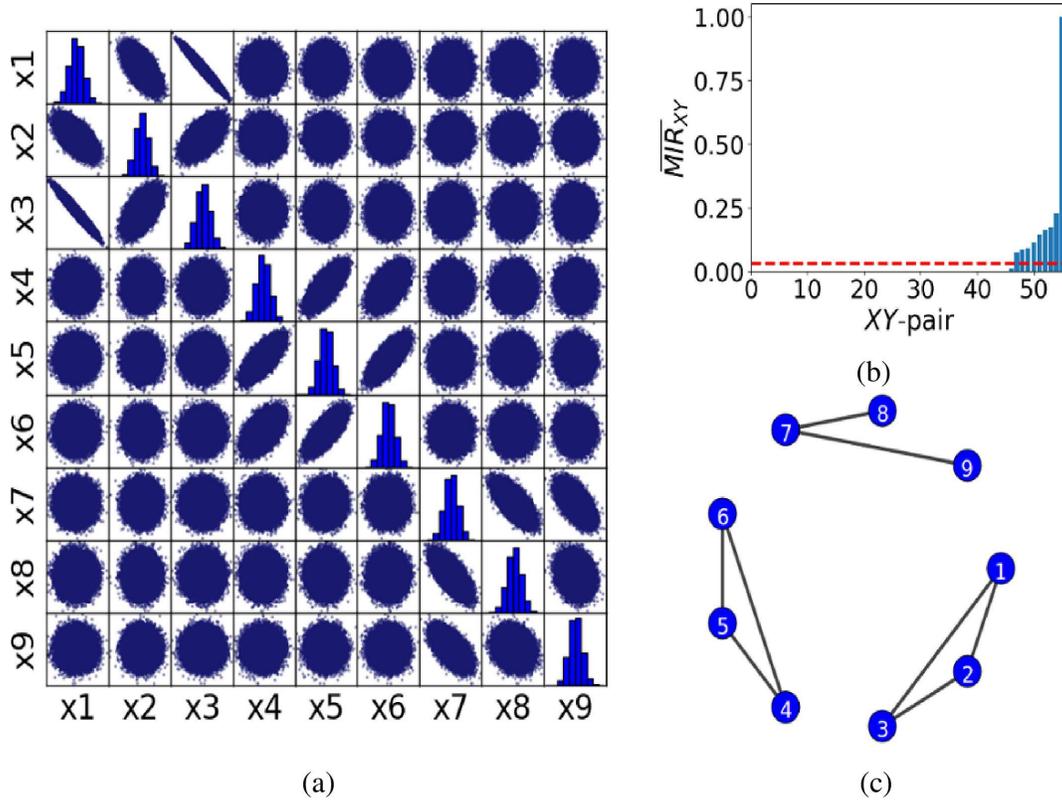}
 \caption{\textbf{Application of the proposed method to correlated normal variates data.} Panel (a) shows the scatter matrix of nine data sets split into three groups (first group: x1,$\ldots$,x3, second group: x4,$\ldots$,x6 and third group: x7,$\ldots$,x9). Each group consists of three correlated normal variates with zero correlation among the groups. Fig~\ref{cor_fig}a shows the scatter matrix of the three groups of data (x1,$\ldots$,x9) . The circular pattern indicates that the two nodes are independent (or weakly correlated), whereas an elongated one shows strong correlation, either positive or negative depending on the slope. Panel (b) is the ordered  $\overline{\mbox{MIR}}_{XY}$ for the correlated variates. The red dash line corresponds to $\tau\approx 0.06$.  Panel (c) shows the successfully inferred networ resulting from (b).}\label{cor_fig} 
\end{figure}

To demonstrate this, we generated three groups of correlated normal variates data. Each group $i=1,2,3$ consists of three correlated normal variates specified by a covariance matrix $\Sigma_i$ and the three groups are uncorrelated with each other. Fig~\ref{cor_fig}a shows the scatter matrix of the three groups of data (first group: x1,$\ldots$,x3, second group: x4,$\ldots$,x6 and third group: x7,$\ldots$,x9) with covariance matrices
$$ \Sigma_1 =
\begin{bmatrix}
  3.40& -2.75& -2.00 \\
 -2.75& 5.50& 1.50 \\
 -2.00& 1.50& 1.25
\end{bmatrix}\hspace{-0.1cm},
\;
\Sigma_2 =
\begin{bmatrix}
 1.0& 0.5& 0.3\\
 0.5& 0.5& 0.3\\
 0.3& 0.3& 0.3 
\end{bmatrix}\hspace{-0.1cm},
\;
\Sigma_3 =
\begin{bmatrix}
 1.40& -2.75& -2.00\\
 -2.75& 5.50& -1.00\\
  -2.00& -1.00& 3.25
 \end{bmatrix}.
$$
A circular pattern for a pair of data sets indicates they are independent (or weakly correlated), whereas an elongated one strong correlation among them, either positive or negative depending on the orientation. For example, in Fig~\ref{cor_fig}a data sets x8 and x9 are weakly correlated and thus one would not expect to see a connection between them in the inferred network. In contrast, since data sets x1 and x3 are strongly anti-correlated, one would expect to see a connection in the inferred network.

This is a case where there is a clear distinction between correlated and non correlated pairs as the data have been constructed as such. This can then be exploited further to set the threshold $\tau$ to identify connectivity in terms of the correlated pairs. Particularly, we set $\tau$ for $\overline{\mbox{MIR}}_{XY}$ so as to depict all correlated pairs of nodes in the data. The results in Fig~\ref{cor_fig}c show that $\overline{\mbox{MIR}}_{XY}$ can be used to infer successfully the network from correlated normal variates. One can see that the data are successfully classified into three distinct groups, and that, the connection between nodes x8 and x9 is missing as they are weakly correlated and the connection between nodes x1 and x3 is present as they are strongly anti-correlated. Since, this pair is also the strongest correlated among all, its $\overline{\mbox{MIR}}_{XY}$ value is also maximal and corresponds to the highest bar in Fig~\ref{cor_fig}b which is equal to 1!

Here, we have shown that the $\overline{\mbox{MIR}}_{XY}$ can depict correctly the number and pairs of correlated data and thus, infer successfully the underlying network structure. Since, global financial data of different economies and countries are also found to be correlated \cite{junior2015dependency}, we will use a similar approach in the next section to infer the network structure of currency exchange rates to USD and stock indices for 14 countries world-wide and the EU.

\FloatBarrier

\section*{Application to financial data}

So far, we have demonstrated the applicability of the method to infer successfully the network structure for artificial data, and we now use it to infer the connectivity in networks of financial-markets data.

Particularly, we have applied the proposed methodology to infer the financial relations among 14 countries world-wide and the EU using the currency exchange rates to US Dollar (USD) and stock indices. The information for the local currencies and stock indices for the 14 countries and the EU are shown in Table \ref{tab:countries}. The data used are daily exchange rates of the local currencies to USD, ranging from January 2000 to August 2016 (taken from the Datastream, Thomson Reuters, database) and stock-indices data, ranging from January 2000 to December 2016 (taken from Bloomberg). We have transformed the daily data points $p_t$ from exchange rates and stock indices to log-return values $r_t = \ln (p_t/p_{t-1})$ where $t$ is the index of the data point in the time-series, as this is a common assumption and practise in Quantitative Finance \cite{hudson2015calculating}.

\begin{table}[!ht]
 \centering
 \begin{tabular}{clll}
  \hline
  Label & Countries/Economies & Local currencies & Stock indices\\
  \hline
  1 & JPN - Japan & Yen & Nikkei 225\\
  2 & EUR - European Union & Euro & Euro Stoxx 50\\
  3 & CAN - Canada & Canadian Dollar &S\&P/TSX Composite Index\\
  4 & TWN - Taiwan & New Taiwan Dollar & TSEC weighted Index \\
  5 & CHE - Switzerland & Swiss Franc & SMI Index\\
  6 & IND - India & Rupee & Bombay BSE 30 \\
  7 & KOR - South Korea & Won & Seoul Composite KS11\\
  8 & BRA - Brazil & Brazilian Real & Bovespa \\
  9 & MEX - Mexico & Mexican Peso & MXX Bolsa Index\\
  10 & NOR - Norway & Norwegian Krone & Oslo OBX Index\\
  11 & SWE - Sweden & Swedish Krona & OMX Stockholm 30 Index\\ 
  12 & SGP - Singapore & Singaporean Dollar & Straits Times Index\\
  13 & ZAF - South Africa & Rand & FTSE/JSE All-Share Index \\
  14 & THA - Thailand & Thai Baht & SET Index\\
  15 & DNK - Denmark & Danish Krone & OMX Copenhagen 20\\
  \hline
 \end{tabular}
 \caption{The local currencies and stock-indices information for the 14 countries and the EU. The first column is the node labels, the second the corresponding 14 countries and the EU (second row), the third their local currency exchange rates with USD and the fourth, the stock indices. Notice that the first column is the node label seen in the networks in Fig~\ref{finance_fig}e and \ref{finance_fig}f.}
 \label{tab:countries}
\end{table}

\begin{figure}[!ht]
  \centering
 \includegraphics[width=0.8\textwidth]{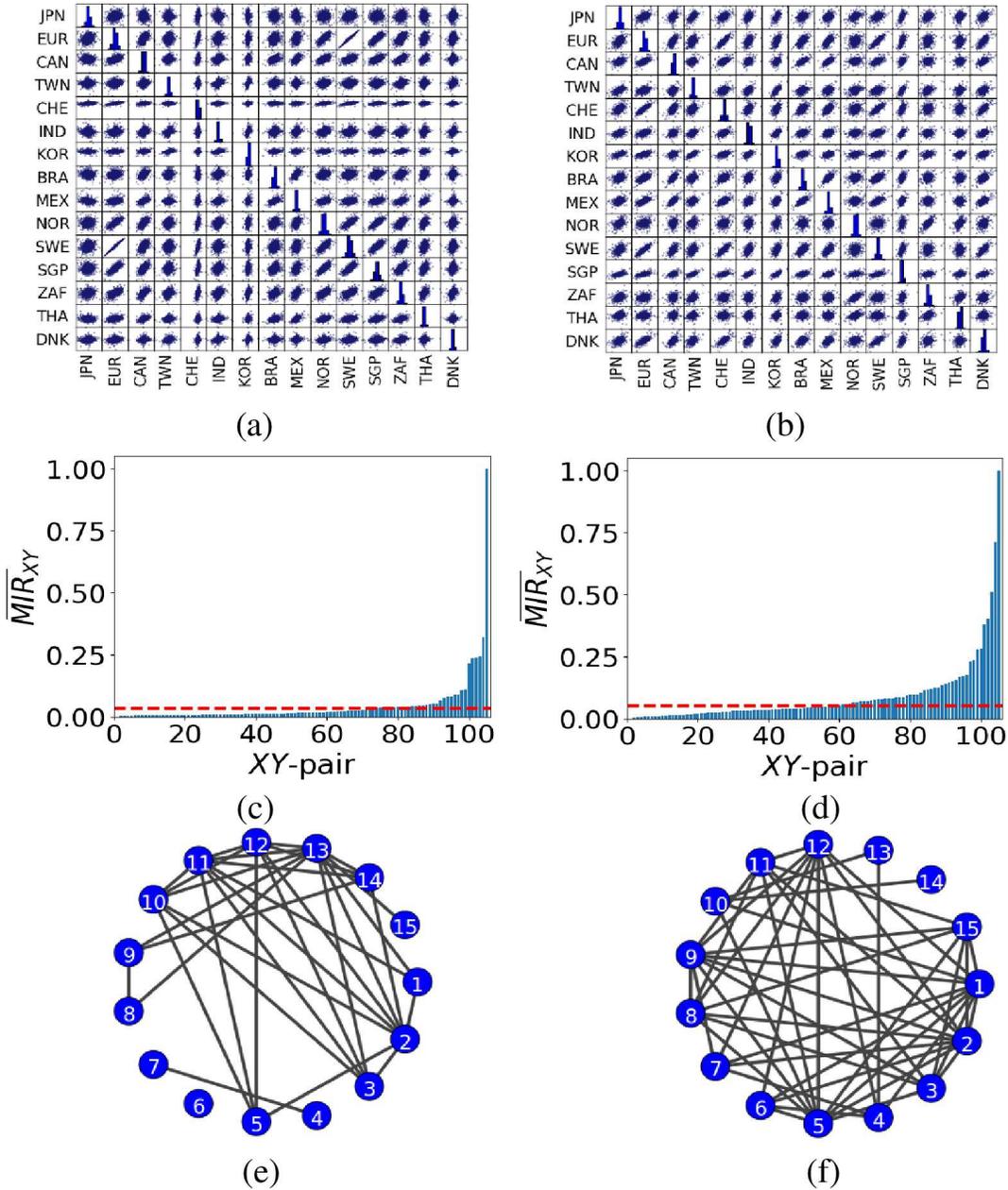}
\caption{\textbf{The scatter plots and inferred networks for the financial-markets data.} Panels (a), (c) and (e) show the scatter matrix, ordered $\overline{\text{MIR}}_{XY}$  values and network of the 15 currency exchange rates. Panels (b), (d) and (f) similarly for the 15 stock indices. The red dash line in (c) corresponds to $\tau \approx 0.03$ and in (d) to $\tau\approx 0.05$. Note that in panels (e) and (f), the nodes represent the currency exchange rates and stock indices for the 15 economies, respectively, as in Table \ref{tab:countries}.}
\label{finance_fig}
\end{figure}

Fig~\ref{finance_fig}a and \ref{finance_fig}b present the scatter plots for the daily log-returns of the exchange rates and stock-indices based on the 15 economies, respectively. In both scatter plots, the strongest correlated pair is the EU-Sweden (EUR-SWE) pair. This strong interrelation is manifested by the highest bars in the ordered $\overline{\mbox{MIR}}_{XY}$ plots in Fig~\ref{finance_fig}c and \ref{finance_fig}d. The second highest, and rest of the bars in Fig~\ref{finance_fig}c do not correspond to the same pairs in Fig~\ref{finance_fig}d, indicating that the connectivity in the two inferred networks for the exchange rates and indices will be different.

To infer the networks from the $\overline{\mbox{MIR}}_{XY}$ plots, we need to set the thresholds appropriately in Fig~\ref{finance_fig}c and \ref{finance_fig}d. Since for the financial data, there is no clear-cut distinction between correlated and non correlated pairs (see Fig~\ref{finance_fig}a and \ref{finance_fig}b), we use the same idea as for the indirect information exchange data and introduce an additional pair of \textit{unidirectional} interacting nodes with chaotic logistic map dynamics to the financial data nodes. The pair is the same for both the forex and stock-indices data and is disconnected from the financial data networks to avoid indirect influences. Again, we assume that any pair of nodes with $\overline{\mbox{MIR}}_{XY}$ bigger than the $\overline{\mbox{MIR}}_{XY}$ of the unidirectional interacting nodes can be regarded as connected. Following this approach, we have found that $\tau\approx0.03$ and $0.05$ for the exchange-rates and stock-indices data, respectively. The difference in the two thresholds comes from the normalisation in Eq. \eqref{MIR_wedge} as the maximum values of $\overline{\mbox{MIR}}_{XY}$ for the forex and stock-indices data are different and they are representing information exchange in different networks. Subsequently, in Fig~\ref{finance_fig}e and \ref{finance_fig}f we present the resulting inferred networks for the daily exchange rates and indices based on the $\overline{\mbox{MIR}}_{XY}$ values for the $105(=15(15-1)/2)$ possible, unique pairs of world-wide economies. Doing so, we have been able to infer most of the weakly and all strongly correlated pairs of economies in both financial data-sets.

To complement our analysis, we have performed a structural analsyis to shed light on the properties of the two inferred networks. Particularly, we have found that both are small-world networks with small-world measures \cite{Basettetal2006,antonopoulos2015brain} of $\sigma\approx5$ for the exchange rates and $\sigma\approx3$ for the stock-indices, respectively. The higher is $\sigma$ from unity, the better it displays the small-world property, with values of $\sigma<1$ indicating a random network. In our case, we find that the stock-indices network is closer to a random network than the exchange-rates network. We have also found that the exchange rates network is a dissasortative (mixing by degree) network (with coefficient of assortativity $r\approx-0.17$) whereas the stock-indices network is assortative with $r\approx0.1$. Assortative mixing by degree is the tendency of nodes with high degree to connect to others with high degree, and similarly for low degree, whereas dissasortative mixing by degree is the tendency of nodes with high degree to connect to other nodes with low degree \cite{2003PhRvE67b6126N}. This is a qualitative difference between the two networks as economies well-connected in terms of their exchange rates prefer to connect with economies less well-connected in terms of the exchange rates as opposed to the stock-indices behaviour where the preference is toward the well connected nodes! Interestingly, our study also reveals that there are 32 bidirectional connections with non-trivial information exchange in the exchange-rates network and 49 in the stock-indices network. Moreover, both networks have a relatively small modularity (i.e. $Q\approx0.12$ and $Q\approx0.14$ for the exchange rates and stock indices, repsectively) indicating that the strength of division of both networks into modules is small, meaning that both networks have sparse connections between the nodes within modules and denser connections between nodes in different modules.

\FloatBarrier

The results in Fig~\ref{finance_fig}e show that the Indian Rupee is not connected to any currency because it is a non-convertible currency, also known as a blocked currency, where the currency cannot be freely traded on the international exchange market, generally as a result of government restrictions. A similar pattern is evident in other non-convertible currencies in this study, namely South Korean Won, Taiwan Dollar and Brazilian Real. The results show that the South Korean Won has no connection with other currencies, except with regional neighbour's currency, Taiwan Dollar. Emerging market currencies, especially those in the same region, tend to mirror each other \cite{kritzer2013forex}. The Brazilian Real and the Mexican Peso have come to epitomise this kind of relationship. Other than Mexican Peso, the Brazilian Real is also connected to South Africa Rand, a fully convertible currency with excellent connection with most currencies around the world.

Our results also confirm that the Euro is the predominant currency in Europe, including countries not in the European Union such as Norway and Switzerland. Two of the three Nordic currencies under study, namely the Swedish Krone and Norwegian Krone, and the Swiss Franc form a network of connections among other and the Euro. The Euro-Swedish Krone pair has the highest $\overline{\mbox{MIR}}_{XY}$ value, close to 1. All Nordic countries considered in this study and Switzerland, are nonetheless intrinsically linked to the European socio-economic and political situations. 

Another interesting finding is that, the Danish Krone, the only Nordic currency evaluated in this study, is not connected with the Euro, although Denmark is a member of the European Union. One reason that can explain this finding could be that Danish Krone is pegged to the Euro, so there is no currency risk that may be affected by economic, monetary or political developments in Europe. In effect, this meant that there are no bidirectional connections in the cases with pegged currency regimes. This can be appreciated in view of Brexit, where the Euro, Swedish Krone, Norwegian Krone and Swiss Franc are slumped against most other currencies, while the Danish Krone, which is pegged to the Euro, is strengthening post-Brexit \cite{wienberg_levring_2016}.

Moreover, our results confirm past studies \cite{wu1998dynamic, bhattacharyya2004integration} that most of the world's major stock indices are integrated into international markets. Fig~\ref{finance_fig}f shows that some of the largest stock-market indices (by market capitalisation) under study, i.e., Nikkei 225 (Japan), Euro Stoxx 50 (European Union) and SMI Index (Switzerland) are connected with most other stock indices. Euro Stoxx 50 and OMX Stockholm 30 Index (Sweden) pair of stock indices has the closest $\overline{\mbox{MIR}}_{XY}$ value to 1 and hence, share the largest rate of information exchange. The Straits Time Index (Singapore) is also well connected with other stock indices internationally, because it is one of the world's most diversified benchmark indices with a mix of stocks that are both domestic and globally focused. The results also show that the SET Index (Thailand) is only connected with one stock index. This is not surprising given that a study by \cite{valadkhani2008dynamic} found no cointegration between stock prices indices of Thailand and its major trading partners.

These networks of currency exchange-rates and stock-indices are relevant for risk managers to use as an investment strategy employed in portfolio management as they give an indication of the mix of assets to hold in order to form a well-diversified portfolio.

\section*{Conclusions}
In this paper, we used the normalised Mutual Information Rate to infer the network structure in artificial and financial data ranging from 2000 to 2016. Particularly, we showed how the underlying network connectivity among the nodes of financial time-series data, such as foreign currency exchange-rates and stock-indices can be inferred. We first demonstrated the applicability of the method by applying it to artificial data from chaotic dynamics and to cases of correlated normal variates. Our results for the artificial data showed that the method can be used to successfully infer the underlying network structures with only using the data recorded from the coupled dynamics assuming no previous knowledge of the adjacency matrices but only to estimate the percentage of successful network inference.

We then applied the method to infer the underlying connectivity structure of currency exchange rates and stock market indices for the economies of 14 worldwide countries and the European Union and, performed an analysis for both networks to identify their structural properties. Our results showed that both are small-world networks with the stock-indices network being assortative by mixing degree and closer to a random network than the exchange-rates network, which was found to be dissasortative by mixing degree. This is a qualitative difference between the two networks as well-connected economies in terms of their exchange rates prefer to connect with economies less well-connected in terms of the exchange rates as opposed to the stock-indices behaviour where the preference is toward the well connected nodes! Interestingly, our study also revealed that both networks have a relatively small modularity, suggesting that both networks have sparse connections between the nodes within modules and denser connections between nodes in different modules.

Our analysis showed that the normalised Mutual Information Rate is a mathematical method that can be used to infer the network structure in complex systems. Since variables in complex systems are not purely random, it is a method to estimate the amount of information exchanged among nodes in systems such as financial markets and identify their connectivity. In our study, the method allowed us to infer two networks and how the world-wide economies are connected to each other and the structural properties of the inferred networks. The currency exchange-rates and stock-indices networks are relevant for risk managers to use as an investment strategy employed in portfolio management as they give an indication of the mix of assets to hold in order to form a well-diversified portfolio.


\end{document}